\documentclass{article}
\usepackage{spconf,amsmath,graphicx,physics,amsfonts}
\usepackage{booktabs, subfig}
\usepackage{comment}
\usepackage[multiple]{footmisc}

\title{A Unified one-shot prosody and speaker conversion system with self-supervised discrete speech units}

\name{Li-Wei Chen, Shinji Watanabe, Alexander Rudnicky}
\address{Language Technologies Institute, Carnegie Mellon University}

\begin{document}
\ninept
\maketitle
\begin{abstract}
We present a unified system to realize one-shot voice conversion (VC) on the pitch, rhythm, and speaker attributes.
Existing works generally ignore the correlation between prosody and language content, leading to the degradation of naturalness in converted speech.
Additionally, the lack of proper language features prevents these systems from accurately preserving language content after conversion. 
To address these issues, we devise a cascaded modular system leveraging self-supervised discrete speech units as language representation.
These discrete units provide duration information essential for rhythm modeling.
Our system first extracts utterance-level prosody and speaker representations from the raw waveform.
Given the prosody representation, a prosody predictor estimates pitch, energy, and duration for each discrete unit in the utterance.
A synthesizer further reconstructs speech based on the predicted prosody, speaker representation, and discrete units.
Experiments show that our system outperforms previous approaches in naturalness, intelligibility, speaker transferability, and prosody transferability.
Code and samples are publicly available.\footnote{https://github.com/b04901014/UUVC}
\end{abstract}
\section{Introduction}
\label{sec:intro}
Human speech carries different aspects of information, including prosody, speaker traits, and language content.
The objective of voice conversion (VC) is to control individual speech attributes with language content unchanged.
In this paper, we focus on the conversion of three main attributes: pitch-energy\footnote{Here we use the term ``pitch-energy''  to refer to only the pitch and energy variations, which is one aspect of prosody.}, speaker traits, and rhythm.

One-shot voice conversion is challenging as the model can only access source and target speech without speaker identities given.
Existing works mostly learned a speaker encoder jointly to isolate speaker information from prosody and language content.
AutoVC~\cite{AUTOVC} attempted to disentangle speaker traits from language by a carefully designed autoencoder.
To separate speaker timbre from prosody, works~\cite{AUTOVCF0} provided pitch contours explicitly to the system.
Several works~\cite{liu19c_interspeech,VQVC+} further improved content separation by learning representations with vector quantization.
Additionally, VQMIVC~\cite{VQMIVC} proposed to minimize mutual information between content, speaker, and pitch representations for better disentanglement.
The above methods, however, focused largely on speaker conversion.
In applications such as emotion style transfer, separate control for prosody is desirable.
Several works built upon AutoVC attempted to control prosodic attributes of speech.
AutoPST~\cite{AUTOPST} modeled rhythm by similarity-based re-sampling.
SpeechSplit~\cite{SpeechSplit,SpeechSplit2} achieved rhythm and pitch conversion with multiple carefully designed autoencoders.
Leveraging these works, SRDVC~\cite{SRDVC} presented a unified one-shot VC system that allows control over both prosody and speaker attributes.

Despite their success, we found that improvements could be made.
Previous approaches often suffered from intelligibility degradation after conversion due to the lack of disentangled language representations.
To address this, recent approaches began to explore self-supervised speech representation~\cite{wav2vec2,wav2vec} (S3R) as a source of language information.
However, these works~\cite{NANSY,s3prl-vc} generally focus on continuous S3R and are limited to speaker conversion.
In contrast, we explore the use of discrete self-supervised speech units on both speaker and prosody conversion.
Compared to continuous S3R, these discrete units formed from clustering naturally encode duration via repeated tokens, which is crucial for rhythm modeling.

Furthermore, we adopt a different modeling approach for prosodic features.
In SRDVC, the pitch representation is directly extracted from a given pitch contour without explicit access to language information.
However, as prosody is correlated with language~\cite{doi:10.1177/002383099704000203}, it causes naturalness degradation of prosody-converted samples (see Section~\ref{ssec:res-pro}).
Energy and duration, although important prosodic features, are also not explicitly modeled.
To address these issues, we propose a cascaded modular system that leverages discrete speech units for language information.
First, our system extracts prosody and speaker representations from the raw waveform.
Given the prosody representation, a prosody predictor estimates the pitch, energy, and duration of each speech unit.
In combination with the predicted prosody, a synthesizer reconstructs speech based on speaker representation and discrete units.
Empirical results demonstrate that our system outperforms previous approaches in intelligibility, naturalness, speaker and prosody transferability.

\section{Method}
\begin{figure*}[!tb]
    \centering
    \includegraphics[width=\linewidth]{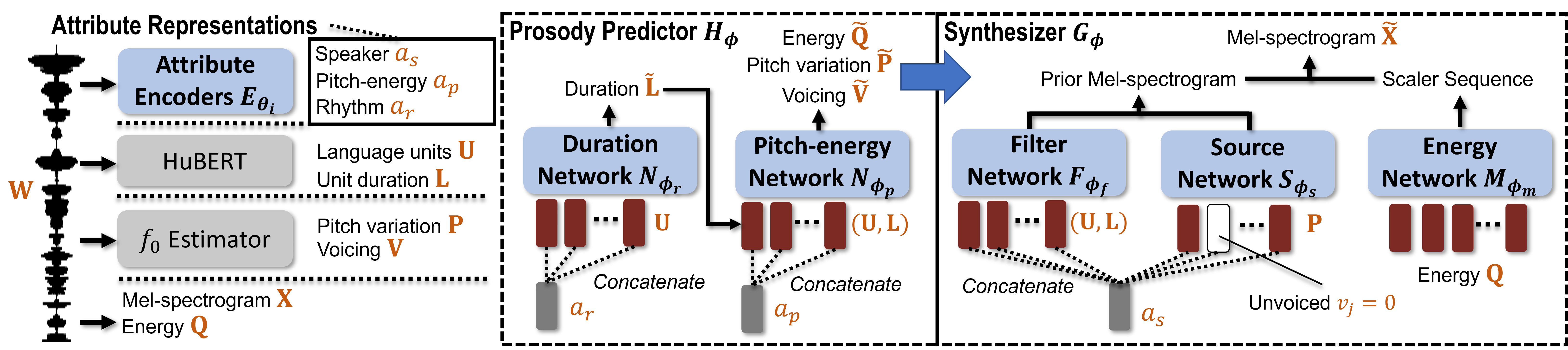}
    \caption{Overview of our system. Attribute Encoders $E_{\theta_i}$ extract attribute representations $\vb{a}_r,\vb{a}_p,\vb{a}_s$. Prosody predictor $H_\phi$ estimates prosodic features based on $\vb{a}_r$ and $\vb{a}_p$. Synthesizer $G_\phi$ reconstructs speech from $\vb{a}_s$ and the estimated prosody.}
    \label{fig:system}
\end{figure*}

\subsection{Problem Formulation}
Figure~\ref{fig:system} presents an overview of the system.
We now describe its corresponding formulation and provide details for each component.
We use the notation $(\cdot)_{i=1}^{I}$ to denote a sequence of length $I$ and $\{\cdot\}_{i=1}^{I}$ to denote a set of $I$ elements.
Given a speech waveform $\vb{W}=(w_t\in \mathbb{R})_{t=1}^T$ and its log-scale mel-spectrogram $\vb{X} = (\vb{x}_n\in \mathbb{R}^{d_x})_{n=1}^N$, we aim to extract different speech attribute representations $\vb{a}_i \in \mathbb{R}^{d_a}$ from $\vb{W}$, and re-synthesis $\vb{X}$ from $\vb{a}_i$.
$T$ and $N$ are the lengths of the waveform and log mel-spectrogram.
$d_x$ is the number of mel-frequency bands, and $d_a$ is a hyper-parameter we choose as the dimension of all $\vb{a}_i$.
We focus on 3 attributes: $i \in \{p, r, s\}$ corresponding to pitch-energy ($\vb{a}_p$), rhythm ($\vb{a}_r$), and speaker ($\vb{a}_s$).
We define speech units in $\vb{W}$ as $\vb{U} = (u_k\in \mathcal{U})_{k=1}^K$, and corresponding duration as $\vb{L} = (l_k\in \mathbb{N})_{k=1}^K$ ($l_k$ is the number of frames each $u_k$ spans).
$K$ is the total number of units for a given utterance, and $\mathcal{U}$ is the set of all possible speech units.
We use $(\vb{U}, \vb{L})$ to denote a new unit sequence formed with duplicating each $u_k$ by $l_k$ across time.
We use learnable embeddings to represent each speech unit in $\mathcal{U}$ as a dense vector.

We further introduce three prosodic features $\vb{P}, \vb{V}, \vb{Q}$ that can be directly inferred from the waveform $\vb{W}$.
The pitch variation sequence $\vb{P}=(p_j \in \mathbb{R})_{j=1}^J$ is mean-normalized pitch in Hertz.
For the voicing sequence $\vb{V} = (v_j \in \{0, 1\})_{j=1}^J$, 0 represents unvoiced and 1 represents voiced.
$J$ is the total number of frames.
We can obtain $\vb{P}$ and $\vb{V}$ from pitch estimator such as CREPE~\cite{CREPE}.
We obtain energy $\vb{Q} = (q_n\in \mathbb{R})_{n=1}^{N}$ from the frame-wise $\mathcal{L}_2$ norm of the linear spectrogram (the same resolution as $\vb{X}$).
Given these, our framework can be initially represented by:
\begin{gather}
    \vb{a}_i = E_{\theta_i}(\vb{W}) \label{eq:attribute_encoders}\\
    \{\Tilde{\vb{P}}, \Tilde{\vb{V}}, \Tilde{\vb{Q}}, \Tilde{\vb{L}}\} = H_{\phi}(\vb{U}, \vb{a}_p, \vb{a}_r) \label{eq:extract}\\
    \Tilde{\vb{X}} = G_\phi(\Tilde{\vb{P}}, \Tilde{\vb{V}}, \Tilde{\vb{Q}}, \vb{U}, \Tilde{\vb{L}}, \vb{a}_s) \label{eq:1}
\end{gather}
where $E_{\theta_i}$ is the attribute encoder for each speech attribute, as shown in the upper-left part of Figure~\ref{fig:system}.
$H_{\phi}$ is our learnable prosody predictor used to estimate prosodic features given units $\vb{U}$ and prosodic attributes $\vb{a}_r, \vb{a}_p$ (corresponding to the middle part of Figure~\ref{fig:system}).
We use $\Tilde{\vb{P}}, \Tilde{\vb{V}}, \Tilde{\vb{Q}}, \Tilde{\vb{L}}$ to denote the network's estimation of $\vb{P}, \vb{V}, \vb{Q}, \vb{L}$.
$G_\phi$ is the synthesizer that reconstructs the speech $\vb{X}$ given the units $\vb{U}$, the predicted prosodic features, and the speaker attribute representation $\vb{a}_s$.
Our objective is designing $G_\phi$ and $H_\phi$ to make $\vb{a}_i$ (equivalently $E_{\theta_i}$) capture the desired speech attributes.
As a consequence, by changing $\vb{a}_i$ we can manipulate different aspect of $\Tilde{\vb{X}}$.

\subsection{Synthesizer $G_\phi$ and Speaker Attribute $\vb{a}_s$}
\label{ssec:learning_a_s}
Inspired by \cite{NANSY}, we follow the source-filter speech production model~\cite{source-filter} to decompose $G_{\phi}$ in Eq.\ref{eq:1} into the addition of two networks.
We further model the energy separately\footnote{During inference, we pass $\Tilde{\vb{P}}, \Tilde{\vb{V}}, \Tilde{\vb{Q}}, \Tilde{\vb{L}}$ instead for Eq.\ref{eq:2}.}:
\begin{align}
    G_\phi&(\vb{P}, \vb{V}, \vb{Q}, \vb{U}, \vb{L}, \vb{a}_s) \nonumber\\
    &= S_{\phi_s}(\vb{P}, \vb{V}, \vb{a}_s) + F_{\phi_f}\left(\vb{U}, \vb{L}, \vb{a}_{s}\right) \oplus M_{\phi_m}(\vb{Q}) \label{eq:2}
\end{align}
where $S_{\phi_s}, F_{\phi_f}$ are source and filter networks that map the input to the same shape as the mel-spectrogram $\vb{X}$ (sequence of length $N$ with dimension $d_x$). $M_{\phi_m}$ is the energy network that outputs a scalar sequence of length $N$.
We use $\oplus$ to denote the broadcast addition across $d_x$.
We now introduce each module in detail.

\textbf{Filter Network.}
Speaker timbre is characterized by the speaker's static articulator shapes (e.g., vocal tract length).
These articulator shapes further influence the phonation of linguistic units $\vb{U}$.
To model this process, we fuse each embedded linguistic unit in $(\vb{U}, \vb{L})$ with $\vb{a}_s$ before passing it to $F_{\phi_f}$, as illustrated in the middle part of Figure~\ref{fig:system}.
The fusion is done simply with channel-wise concatenation followed by a linear layer.
For the network architecture, we adopted repeated residual blocks which consists of 1d-CNN, ReLU activation, linear layer, residual connection and layer normalization~\cite{layernorm}.
The same structure is also used for all networks $F_{\phi_f},S_{\phi_s},M_{\phi_m},N_{\phi_p},N_{\phi_r}$.
Since the length of $(\vb{U}, \vb{L})$ is different from that of $\vb{X}$ (length $N$), we use nearest interpolation on the output of an intermediate block to align two time sequences.

\textbf{Source Network.}
The source network $S_{\phi_s}$ processes the pitch variation sequence $\vb{P}$ and voicing sequence $\vb{V}$ into an excitation spectrogram.
We first follow a similar procedure in \cite{CREPE} to process pitch.
Specifically, we first define $B$ equal frequency bins $\vb{C}^p = \{c_i=i\ell_{b} + p_{\text{min}}\}_{i=1}^B$.
We choose the minimum normalized pitch $p_{\text{min}} = -250$, and the bin width $\ell_{b}=2.5$.
We then calculate $\vb{b}(p_j) = \{b_{i}(p_j)\}_{i=1}^{B}$, the bin weight of $c_i$ for $p_j$ by applying Gaussian blur:
\begin{gather}
    b_{i}(p_j) = \exp\left(-\frac{(p_j - c_i)^2}{2\sigma^2}\right)\label{eq:bin_weight}
\end{gather}
We choose the blur standard deviation $\sigma = 4$, and $B = 200$.
Finally, we compute the dense representation of $p_j$: $\vb{o}(\vb{b}(p_j), \vb{E}^p)$ by the weighted sum of a set of randomly initialized learnable embeddings $\vb{E}^p = \{\vb{e}^p_i \in \mathbb{R}^{d_e}\}_{i=1}^B$ with the bin weights $\vb{b}(p_j)$:
\begin{gather}
    \vb{o}(\vb{b}(p_j), \vb{E}^p) = \frac{\sum_{i=1}^B b_{i}(p_j)\vb{e}^p_i}{\sum_{i=1}^B b_{i}(p_j)}\label{eq:bin_embed}
\end{gather}
$d_e$ is the dimension of the embedding.
To include voicing information $\vb{V}$, we replace the unvoiced frames ($j$ where $v_j = 0$) of $\vb{o}(\vb{b}(p_j), \vb{E}^p)$ with another learnable embedding before further processing.
Similar to the filter network, we provide $\vb{a}_s$ to each time frame of $\vb{o}(p_j)$.
However, $\vb{a}_s$ is now responsible for recovering the average $f_0$ discarded in mean normalization of $\vb{P}$.
Finally, we add the output of the filter and source network (both the same shape as $\vb{X}$) and term the output as the prior mel-spectrogram.
It carries most of the information $(\vb{P}, \vb{U}, \vb{L}, \vb{a}_s)$ needed to reconstruct the original speech, except for energy.

\textbf{Energy Network.}
Another attribute unrelated to speaker timbre is energy.
While directly adding energy $\vb{Q}$ to the prior mel-spectrogram ($S_{\phi_s} + F_{\phi_f}$) is feasible, this restricts the prior mel-spectrogram to have equal spectral energy across time.
Instead, we train another network $M_{\phi_m}$ to process energy, as shown in Eq.\ref{eq:2} and the right part of Figure~\ref{fig:system}.
We adopt a similar procedure of encoding pitch $\vb{P}$ mentioned in the previous paragraph to encode energy $\vb{Q}$.\footnote{We choose the minimum energy to be 0 and the energy bin width to be 1 for the calculation of energy bins $\vb{C}^q$.}
We denote the corresponding outcome of Eq.\ref{eq:bin_weight} and Eq.\ref{eq:bin_embed} as $b_i(q_n)$ and $\vb{o}(\vb{b}(q_n), \vb{E}^q)$, where $\vb{E}^q = \{\vb{e}^q_i \in \mathbb{R}^{d_e}\}_{i=1}^B$ is the learnable embedding for energy bins.
$M_{\phi_m}$ then maps the encoded energy sequence $\{\vb{o}(\vb{b}(q_n), \vb{E}^q)\}_{n=1}^N$ to a scalar sequence of length $N$.

\subsection{Prosody Predictor $H_\phi$ and Prosodic Attributes $\vb{a}_p,\vb{a}_r$}
\label{ssec:pro}
We now introduce the prosody predictor $H_\phi$ in Eq.\ref{eq:extract}.
We decompose $H_\phi$ into the cascade of duration network $N_{\phi_r}$ and pitch-energy network $N_{\phi_p}$.
We first model rhythm by duration prediction:
\begin{gather}
    \Tilde{\vb{L}} = N_{\phi_r}(\vb{U}, \vb{a}_{r}) \label{eq:rhythm}
\end{gather}
The rhythm representation $\vb{a}_r$ is trained to encode information useful for recovering the duration $\vb{L}$ from unit $\vb{U}$.
For the prediction of pitch, we condition on $(\vb{U}, \vb{L})$ and $\vb{a}_p$:
\begin{gather}
    (\Tilde{\vb{P}}, \Tilde{\vb{V}}, \Tilde{\vb{Q}}) = N_{\phi_p}(\vb{U}, \vb{L}, \vb{a}_{p})\label{eq:pitch}
\end{gather}
We predict $\Tilde{\vb{P}}, \Tilde{\vb{V}}, \Tilde{\vb{Q}}$ jointly with $\vb{a}_p$ as they are highly correlated.
We condition on $(\vb{U}, \vb{L})$ as prosody is correlated with language content~\cite{doi:10.1177/002383099704000203}.
For $\Tilde{\vb{P}}$ and $\Tilde{\vb{Q}}$, instead of predicting the scalar values, we follow \cite{CREPE} to predict bin weights.
We use $\Tilde{\vb{b}}^p_{j},\Tilde{\vb{b}}^q_{n}$ to denote the network estimation of bin weights $\vb{b}(p_j),\vb{b}(q_n)$ calculated by Eq.\ref{eq:bin_weight}.
Note that for the source and energy network, it is sufficient to provide bin weights to calculate encoded pitch and energy with Eq.\ref{eq:bin_embed}.

\subsection{Attribute Encoders $E_{\theta_i}$}
Studies~\cite{Fine-grained-TTS,NANSY} have found Wav2Vec 2.0~\cite{wav2vec2} useful for learning utterance-level representation.
Here we adopt pretrained\footnote{https://huggingface.co/facebook/wav2vec2-base} Wav2vec 2.0 as our attribute encoders $E_{\theta_i}$ in Eq.\ref{eq:attribute_encoders}.
Wav2Vec 2.0 consists of a cascade of CNN feature extractor and transformer encoder blocks.
We only use its CNN feature extractor and the first layer of its transformer encoder.
Specifically, we fixed the CNN feature extractor untrained, and fine-tune the 1-layer transformer separately for each $E_{\theta_i}$.
We then apply global average pooling on the output to collapse the time dimension, followed by a linear layer to obtain $\vb{a}_{i}$.

\subsection{Optimization}
\label{ssec:opt}
We adopt $\mathcal{L}_1$ loss for the reconstruction of $\Tilde{\vb{X}}$ and $\vb{X}$.
We further apply commonly used adversarial loss to prevent over-smoothness of $\Tilde{\vb{X}}$. 
We use least-squared adversarial loss~\cite{Mao_2017_ICCV} with the discriminator architecture following \cite{NANSY}.
To predict prosodic features with $H_\phi$, we minimize the binary cross entropy (BCE) between $\vb{V}$ and $\Tilde{\vb{V}}$.
We follow \cite{FASTSPEECH} to use mean squared error (MSE) as loss function on log-scale duration for $\Tilde{\vb{L}}$.
For $\Tilde{\vb{P}}$ and $\Tilde{\vb{Q}}$, we follow \cite{CREPE} to minimize BCE between predicted and ground truth bin weights (e.g. $\Tilde{\vb{b}}^p_{j}$ and $\vb{b}(p_j)$).
During training, we use the ground truth $\vb{L},\vb{V}$ as the input to all networks.
During inference, we pass the predicted $\Tilde{\vb{L}},\Tilde{\vb{V}}$.

\textbf{Joint optimization.}
During training, a reasonable choice is to use the ground truth $\vb{P},\vb{Q}$ as the input to the synthesizer $G_\phi$ in Eq.\ref{eq:1}.
However, this leaves no training signal between the synthesizer $G_\phi$ and the prosody predictor $H_\phi$ in Eq.\ref{eq:extract}.
One would imagine that the reconstruction and adversarial loss of $\Tilde{\vb{X}}$ can guide the pitch-energy network $N_{\phi_p}$ to generate a perceptually natural pitch and energy contour.
One straightforward solution is to pass the estimated pitch ($\Tilde{\vb{b}}^p_{j}$) and energy ($\Tilde{\vb{b}}^q_{n}$) to $G_\phi$.
Empirically, we found that this leads to $G_\phi$ ignoring the given prosodic information.
We attribute this to the poor estimation in the early stage of training, which discourages $G_\phi$ to consider prosodic information.
We instead pass the average of predicted and ground truth bin weights (e.g.,  $0.5(\vb{b}(p_j) + \Tilde{\vb{b}}^p_{j})$) to $G_\phi$ during training.
Additionally, we minimize the MSE loss between the predicted and ground truth encoded pitch ($\vb{o}(\vb{b}(p_j), \vb{E}^p)$ and  $\vb{o}(\Tilde{\vb{b}}^p_{j}, \vb{E}^p)$).
This explicitly encourages $\vb{E}^p$ to have consistent encoding between $\vb{b}(p_j)$ and $\Tilde{\vb{b}}^p_{j}$.
The same loss is applied to energy.

\section{Experimental Setup}
We use 16 blocks for $F_{\phi_f},S_{\phi_s}$, 4 blocks for $M_{\phi_m}$, 2 blocks for $N_{\phi_r}$ and 6 blocks for $N_{\phi_p}$.
The block is the residual block we introduced in Section~\ref{ssec:learning_a_s}.
We apply k-means clustering on the final layer of HuBERT~\cite{HuBERT} with 200 clusters ($|\mathcal{U}|=200$) as our self-supervised speech units.
We use the textless-lib~\cite{textless-lib} to extract pitch $\vb{P}$, voicing $\vb{V}$ and the HuBERT units $(\vb{U}, \vb{L})$.
We follow \cite{kharitonov-etal-2022-text} to deduplicate consecutive HuBERT units as $\vb{U}$, and the number of duplicated frames form $\vb{L}$.
We adopt pretrained HiFi-GAN~\cite{HIFI} for mel-spectrogram inversions.
We include detailed implementation in the released code.
Given a source speech and a target speech, we evaluate our system with two settings: speaker conversion (transfer $\vb{a}_s$ from target) and prosody conversion (transfer $\vb{a}_r,\vb{a}_p$ from target).

\textbf{Datasets.}
We follow previous approaches~\cite{AUTOVC,VQMIVC,SRDVC} to train on the VCTK corpus~\cite{VCTK}.
VCTK consists of reading English with 400 sentences each from 110 speakers.
We randomly sample 5\% utterances for validation and the rest for training.
To fully evaluate the one-shot capability, we test on utterances from LibriTTS~\cite{LibriTTS}.
LibriTTS is another reading English corpus with larger speaker and language content coverage (over 2000 speakers), providing a more unbiased evaluation.
We randomly sample 1000 utterances each as the source and target speech respectively for both speaker and prosody transfer.

\textbf{Comparing Methods.}
AutoVC~\cite{AUTOVC}, SRDVC~\cite{SRDVC} are chosen as competing methods for speaker conversion.
We used the officially released checkpoints and HiFi-GAN vocoder for these methods.
For prosody conversion, we compare our method with SRDVC.
We further evaluate two variants of our system.
First, we replace the pretrained Wav2Vec 2.0 with random initialization (\textit{-W2V2} in the tables).
Second, we evaluate the system without the joint optimization losses we introduced in Section~\ref{ssec:opt} (\textit{-Joint. Opt.} in the tables).

\textbf{Metrics.}
We report the character error rate (CER) of the syntheses to measure intelligibility~\cite{ESPNET-TTS}.
We use google ASR API for the transcription.
For objective metrics of speaker conversion, we report the cosine similarity between speaker embeddings of target and converted speech.
We term this speaker embedding similarity (SES); its range is between $[-1, 1]$.
Speaker embeddings are extracted with a pretrained speaker identification network\footnote{https://github.com/pyannote/pyannote-audio}.
Similarly, we extract emotion embeddings from a pretrained dimensional emotion classifier~\cite{wagner2022dawn}.
We again report the cosine similarity of emotion embeddings (EES).
For subjective measures, human evaluations are conducted via Amazon Mechanical Turk.
We randomly sampled 25 utterances from each method and assigned them to 10 workers.
We report both the 5-scale mean opinion score (MOS) and 95\% confidence interval (CI) on naturalness, speaker and prosody similarity.

\section{Results}
\label{sec:result}
\begin{table}[!tb]
    \centering
    \caption{Evaluation results for \textbf{unseen speaker transfer}.
    The right columns are naturalness and speaker similarity MOS with 95\% CI.
 \textit{GT} stands for the ground truth speech.}
    \label{tab:speaker}
    \begin{tabular}{l|c c|c c}
        \toprule
        Method & CER $\downarrow$  & SES $\uparrow$ & Naturalness $\uparrow$ & Similarity $\uparrow$ \\
        \midrule
        \textit{GT} & 5.5\% & \textit{n/a} & 3.95 $\pm$ 0.11 & \textit{n/a} \\
        \midrule
        \textit{AutoVC} & 88.4\% & 0.14 & 2.58 $\pm$ 0.17 & 2.62 $\pm$ 0.15 \\
        \textit{SRDVC} & 34.7\% & 0.17 & 3.25 $\pm$ 0.15 & 2.56 $\pm$ 0.14 \\
        \midrule
        \textit{Ours} & 7.5\% & \textbf{0.34} & 3.50 $\pm$ 0.14 & \textbf{2.80} $\pm$ 0.14 \\
        \textit{(-W2V2)} & 7.8\% & 0.27 & 3.47 $\pm$ 0.13 & 2.62 $\pm$ 0.15\\
        \textit{(-Joint Opt.)} & \textbf{7.3\%} & 0.32 & \textbf{3.55} $\pm$ 0.14 & 2.68 $\pm$ 0.14 \\
        \bottomrule
    \end{tabular}
\end{table}

\begin{table}[!tb]
    \centering
    \caption{Average Pearson correlation coefficients (PCC) of pitch and energy between target and speaker-converted speech.}
    \label{tab:disentangle}
    \begin{tabular}{l|c c|c c}
        \toprule
        PCC & \textit{AutoVC} & \textit{SRDVC} & \textit{Ours $(\Tilde{\vb{P}},\Tilde{\vb{Q}})$} & \textit{Ours $(\vb{P},\vb{Q})$}\\
        \midrule
        $\log{f_0}$ & 0.09 & 0.49 & 0.30 & \textbf{0.51} \\
        Energy & 0.05 & 0.82 & 0.78 & \textbf{0.91} \\
        \bottomrule
    \end{tabular}
\end{table}

\subsection{Speaker Conversion}
Table~\ref{tab:speaker} presents our experiment result on unseen speaker transfer.
First, compared to AutoVC and SRDVC, our system achieves much higher intelligibility (lower CER) and naturalness.
Apparently, the discrete self-supervised units provide better language information.
For speaker transferability, our system achieves the highest similarity MOS and SES.
The similarity MOS noticeably drops if we do not use pretrained Wav2Vec 2.0, suggesting that pretrained SSL models can capture more generalizable speaker characteristics.
Without the joint optimization losses, the system obtains distinctively worse speaker similarity.
For the synthesizer $G_\phi$, joint optimization can be interpreted as a means of data augmentation for the given prosody.
We conjecture that this encourages $G_\phi$ to learn more generalized $\vb{a}_s$ for a wider variety of prosody patterns.

\textbf{Disentanglement of speech attributes.}
Table~\ref{tab:disentangle} presents the average Pearson correlation coefficients (PCC) of prosodic features between target and speaker-converted speech.
Higher PCC suggests that the corresponding attribute is less affected following speaker conversion.
We analyze our system under two different conditions: passing reconstructed $\Tilde{\vb{P}}$, $\Tilde{\vb{Q}}$, and passing the ground truth $\vb{P}$, $\vb{Q}$.
With the ground truth $\vb{P}$, $\vb{Q}$, our system achieves comparable PCC on pitch and distinctively higher PCC on energy compared to SRDVC.\footnote{Note that SRDVC also accepts ground truth pitch contour as input.}
This shows the effectiveness of our energy modeling approach to disentangle $\vb{a}_s$ and the energy contour (speaker conversion will not affect energy).
On the other hand, using $\Tilde{\vb{P}}$, $\Tilde{\vb{Q}}$ leads to a lower PCC, suggesting that the pitch-energy reconstruction from $\vb{a}_{r}, \vb{a}_{p}$ is not perfect.
Note that this is both reasonable and desirable; our goal for $\vb{a}_{p}$ is not to memorize the exact pitch (energy) contour but to model the high-level speaking style of the utterance.
The mapping from speaking style to the pitch contour is one-to-many, which naturally results in lower PCC.

\subsection{Prosody Conversion}
\label{ssec:res-pro}
\begin{table}[!tb]
    \centering
    \caption{Evaluation results for \textbf{prosody transfer}. The right columns are naturalness and prosody similarity MOS with 95\% CI.} 
    \label{tab:prosody}
    \begin{tabular}{l|c c |c c}
        \toprule 
        Method & CER $\downarrow$ & EES $\uparrow$ & Naturalness $\uparrow$ & Similarity $\uparrow$ \\
        \midrule
        \textit{GT} & 5.5\% & \textit{n/a} & 3.95 $\pm$ 0.11 & \textit{n/a} \\
        \midrule
        \textit{SRDVC} & 49.9\% & 0.28 & 3.06 $\pm$ 0.16 & 2.58 $\pm$ 0.15\\
        \midrule
        \textit{Ours} & \textbf{8.9\%} & \textbf{0.42} & \textbf{3.49} $\pm$ 0.13 & \textbf{2.69} $\pm$ 0.14\\
        \textit{(-W2V2)} & 10.1\% & 0.35 & 3.39 $\pm$ 0.14 & 2.57 $\pm$ 0.15\\
        \textit{(-Joint Opt.)} & 9.0\% & 0.38 & 3.36 $\pm$ 0.14 & 2.50 $\pm$ 0.14 \\
        \bottomrule
    \end{tabular}
\end{table}
Table~\ref{tab:prosody} shows the result for prosody conversion.
We first observe that SRDVC suffers from much higher CER and lower naturalness MOS compared to its own performance in speaker conversion, suggesting that it failed to generate natural prosody.
We observe that in many samples SRDVC directly transfers the pitch contour from the target speech without considering phonetic content.
Its high similarity MOS (even higher than our ablated versions) but significantly lower naturalness further supports this claim.
On the other hand, compared to speaker conversion, our system shows a small increase in CER and almost no difference in naturalness MOS.
Since our prosody prediction is conditioned on discrete speech units, consistency between language content and prosody can be learned naturally.
Additionally, our system better transfers prosody, judging from the highest EES score and similarity MOS.
Table~\ref{tab:prosody} also shows that joint optimization noticeably increases prosody naturalness and similarity.
It suggests that the adversarial loss and reconstruction loss indeed provide useful guidance to the prosody predictor $H_\phi$.
All measures noticeably degrade without pretrained SSL models, indicating its usefulness for extracting speaking styles.

\subsection{Visualization of Prosody Representations.}
\begin{figure}[!tb]
    \centering
    \includegraphics[width=\linewidth]{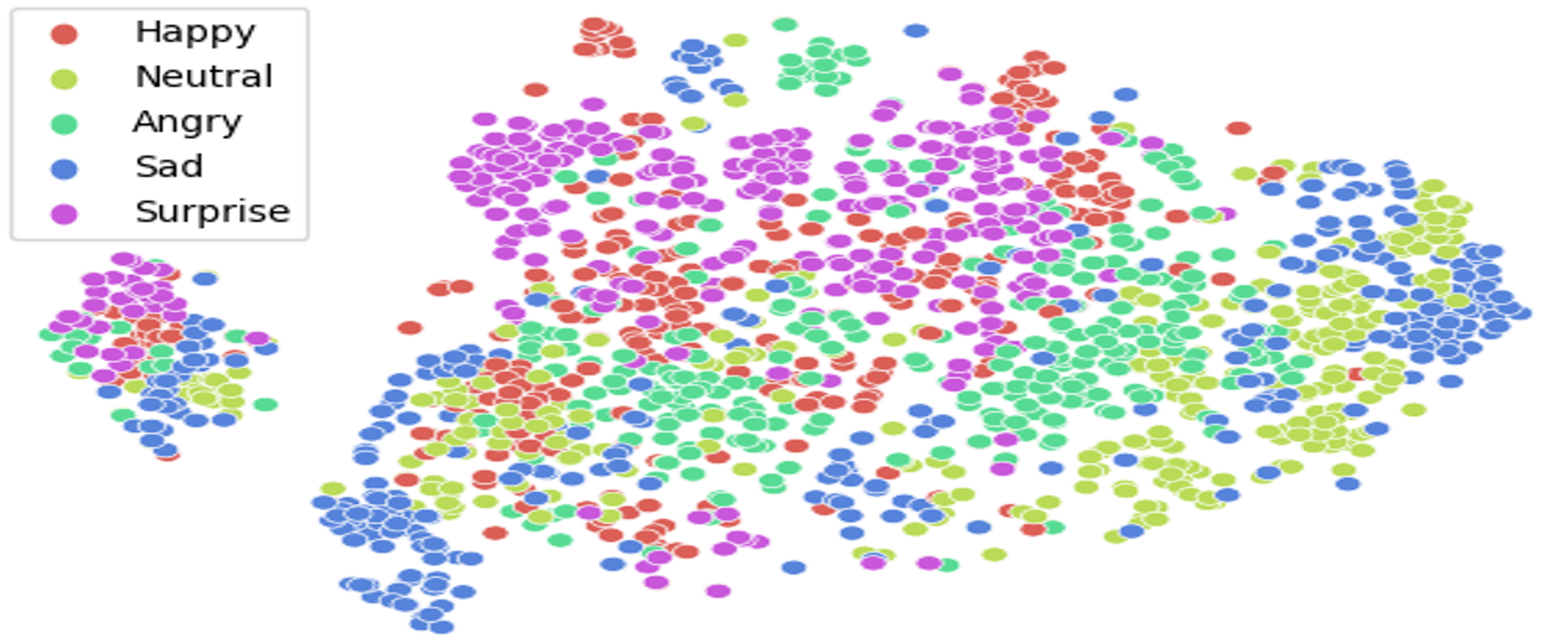}
    \caption{Visualization of $\vb{a}_p$ (pitch-energy) by running t-SNE on ESD.}
    \label{fig:tsne}
\end{figure}
We further visualize $\vb{a}_p$ by running t-SNE~\cite{t-sne} on the Emotional Speech Database (ESD)~\cite{ESD}.
ESD contains 350 English utterances spoken by 10 speakers with 5 emotion categories.
From Figure~\ref{fig:tsne}, we observe that despite being trained on only reading speech (VCTK), $\vb{a}_p$ still forms clusters of emotions.
In particular, sad and neutral mostly covers the lower part while happy and surprise concentrate on the upper part of Figure~\ref{fig:tsne}.
This validates that $\vb{a}_p$ indeed captures high-level speaking style information.

\section{Conclusion}
We describe a unified system for one-shot prosody and speaker conversion trained in an unsupervised manner.
We evaluate the intelligibility, naturalness, speaker and prosody transferability of synthetic speech and show the superior performance of our approach.
Our work potentially benefits various downstream tasks including voice conversion, emotion analysis, speech data augmentation, and expressive speech synthesis.
Based on this work, we intend to extend our system to real-world speech, where background noise and recording environments are additional attributes to consider.
We also plan to investigate the potential downstream applications of learned attribute representations.

\section{Acknowledgements}
We are grateful to Amazon for the support of this research.

\bibliographystyle{IEEEbib}
\bibliography{main}

\end{document}